\documentclass[%
 aip,
 amsmath,amssymb,
 reprint,%
]{revtex4-2}

\usepackage{graphicx, caption}
\usepackage{}
\usepackage{bm}
\usepackage{hyperref}
\usepackage[utf8]{inputenc}
\usepackage[T1]{fontenc}
\usepackage{mathptmx}
\usepackage{amsmath}
\usepackage{setspace} 
\usepackage{xcolor}

\usepackage{float}
\newcommand*{\citen}[1]{%
  \begingroup
    \romannumeral-`\x 
    \setcitestyle{numbers}%
    \cite{#1}%
  \endgroup
}

\begin{document}
\long\def\/*#1*/{}
\preprint{AIP/123-QED}

\title[Elevating zero dimensional global scaling predictions]{Elevating zero dimensional global scaling predictions to self-consistent theory-based simulations}

\author{T. Slendebroek}%
\email{slendebroekt@fusion.gat.com}
\affiliation{Oak Ridge Associated Universities, Oak Ridge, TN 37830, USA}
\author{J. McClenaghan}%
\affiliation{General Atomics, San Diego, California 92121, USA 
}%

\author{O.M. Meneghini}
\affiliation{General Atomics, San Diego, California 92121, USA 
}%

\author{B.C. Lyons}
\affiliation{General Atomics, San Diego, California 92121, USA 
}%

\author{S.P. Smith}
\affiliation{General Atomics, San Diego, California 92121, USA 
}%
\author{T.F. Neiser}
\affiliation{General Atomics, San Diego, California 92121, USA 
}%

\author{N. Shi}
\affiliation{General Atomics, San Diego, California 92121, USA 
}%

\author{J.~Candy}
\affiliation{General Atomics, San Diego, California 92121, USA 
}%
\date{\today}

\begin{abstract}
We have developed an innovative workflow, STEP-0D, within the OMFIT integrated modelling framework. Through systematic validation against the International Tokamak Physics Activity (ITPA) global H-mode confinement database, we demonstrated that STEP-0D, on average, predicts the energy confinement time with a mean relative error (MRE) of less than 19\%. Moreover, this workflow showed promising potential in predicting plasmas for proposed fusion reactors such as ARC, EU-DEMO, and CFETR, indicating moderate H-factors between 0.9 and 1.2. STEP-0D allows theory-based prediction of tokamak scenarios, beginning with zero-dimensional (0D) quantities. The workflow initiates with the PRO-create module, generating physically consistent plasma profiles and equilibrium using the same 0D quantities as the IPB98(y,2) confinement scaling. This sets the starting point for the STEP (Stability, Transport, Equilibrium, and Pedestal) module, which further iterates between theory-based physics models of equilibrium, core transport, and pedestal to yield a self-consistent solution. Given these attributes, STEP-0D not only improves the accuracy of predicting plasma performance but also provides a path towards a novel fusion power plant (FPP) design workflow. When integrated with engineering and costing models within an optimization, this new approach could eliminate the iterative reconciliation between plasma models of varying fidelity. This potential for a more efficient design process underpins STEP-0D's significant contribution to future fusion power plant development.
\end{abstract}

\maketitle


\section{Introduction}
Viable fusion power plants require a burning or ignited plasma core to produce a net positive electrical output. Experimental devices, up to this point, have achieved production of 3 MW of fusion power over the span of 4 seconds \cite{keilhacker1999d}. Recent JET deuterium-tritium experiments have demonstrated enhanced fusion performance, achieving 10 MW for 5 seconds \cite{maslov2022t}, by optimizing the deuterium-tritium ratio, which in turn increases the beam-target fusion rate. Despite these noteworthy accomplishments, the transition from short-lived fusion power bursts to sustainable operation over extended periods is far from straightforward. Operating in a pulsed manner for approximately three hours to steady-state, powered by thermalized fusion reactions with a net electric output, remains an ongoing challenge. Various approaches have been proposed to meet this challenge such as larger size like EU-DEMO \cite{siccinio2022development}, higher magnetic field approaches such as ARC \cite{sorbom2015arc} or advanced plasma operation approaches, such as CAT \cite{buttery2021advanced}. The space of possible tokamak configurations, in terms of engineering and plasma parameters, is vast and expands as advances are made in technology such as the improvements in magnetic field with the use of high temperature superconductors and advances in understanding of the plasma core through high confinement regimes. How do these advances and improvements translate to the performance of the plasma core? 0D systems codes \cite{Chan_2015, reux2015demo} are frequently used to scope the wide and multi-dimensional parameter space of tokamaks. These 0D systems codes predict the performance of the core plasma by relying on scaling laws, notably the IPB98(y,2) confinement scaling \cite{christiansen1992global}. This scaling law uses 8 physically meaningful parameters that describe the plasma and geometry of the tokamak and is derived from a weighted least squares regression of the  International Tokamak Physics Activity (ITPA) global H-mode confinement database as a power law. The use of this power law is robust and accurate for the given data range. However, the parameter space of a fusion power plant (FPP) lies outside the database range and thus extrapolations of the scaling law are required which adds uncertainty which is difficult to quantify.

In contrast to extrapolation with scaling laws, predictions based on integrated modeling workflows remain valid provided that the underlying physics-based models are used within their applicable bounds, which extend to the FPP operational domain. For example, the flux-driven transport solver TGYRO \cite{candy2009tokamak} and pedestal stability code EPED \cite{snyder2009development} have been successfully used in integrated modeling workflows for both existing experiments and future burning plasma devices [\citen{rodriguez2020predictions}, \citen{snyder2019high},\citen{knolker2020optimizing}]. Notably, initial performance predictions of burning plasmas in SPARC \cite{creely2020overview} using a reduced model of gyrokinetic transport (TGLF \cite{staebler2020geometry}) have recently been successfully validated using high-fidelity gyrokinetic simulations with CGYRO \cite{candy2016high} in a novel integrated modeling workflow [\citen{rodriguez2022nonlinear}, \citen{rodriguez2022overview}]. Integrated modeling has been used to reach a self consistent 1.5D solution by iterating over core sources, transport, pedestal and equilibrium codes. Predictions of the self-consistent simulations on present day devices have shown good agreement with experimental results [\citen{luda2020integrated}, \citen{hyun_tae_2023}, \citen{staebler2022advances}]. A key aspect of these integrated simulations is the initial conditions from which the iteration process is started. For example, a widely used starting point to predict the performance of existing devices is to begin from a past experimental condition and then modify original actuator settings \cite{garofalo2015compatibility}. Another approach is to start from databases of existing simulations \cite{murakami2011integrated}. Both of these methods require significant effort to finding data either experimental or simulated and are limited in their ability to rapidly explore a wide range of parameters. 

In this paper we present a method for bootstrapping 1.5D stationary simulations of fusion plasmas, starting from the scalar (i.e. 0D) engineering parameters that are commonly used for cross-machine studies and extrapolation to fusion power plants using 0D systems codes (see section I-A). We used this approach to efficiently initialize self-consistent 1.5D simulations (see section I-B). We validate this approach against DIII-D shots (see section II-A) and existing cross-machine databases such as the ITPA global H-mode confinement database \cite{verdoolaege2021updated} from which the IPB98(y,2) confinement scaling \cite{christiansen1992global} is derived (see section II-B). Predictions of proposed fusion reactors were made in section II-C. We also summarize this research and provide perspective on future work (see Section III).

\subsection{Bootstrapping a 1.5D simulation from 0D parameters with PRO-create}

\begin{figure*}
    \centering
    \includegraphics[width=1.0\textwidth]{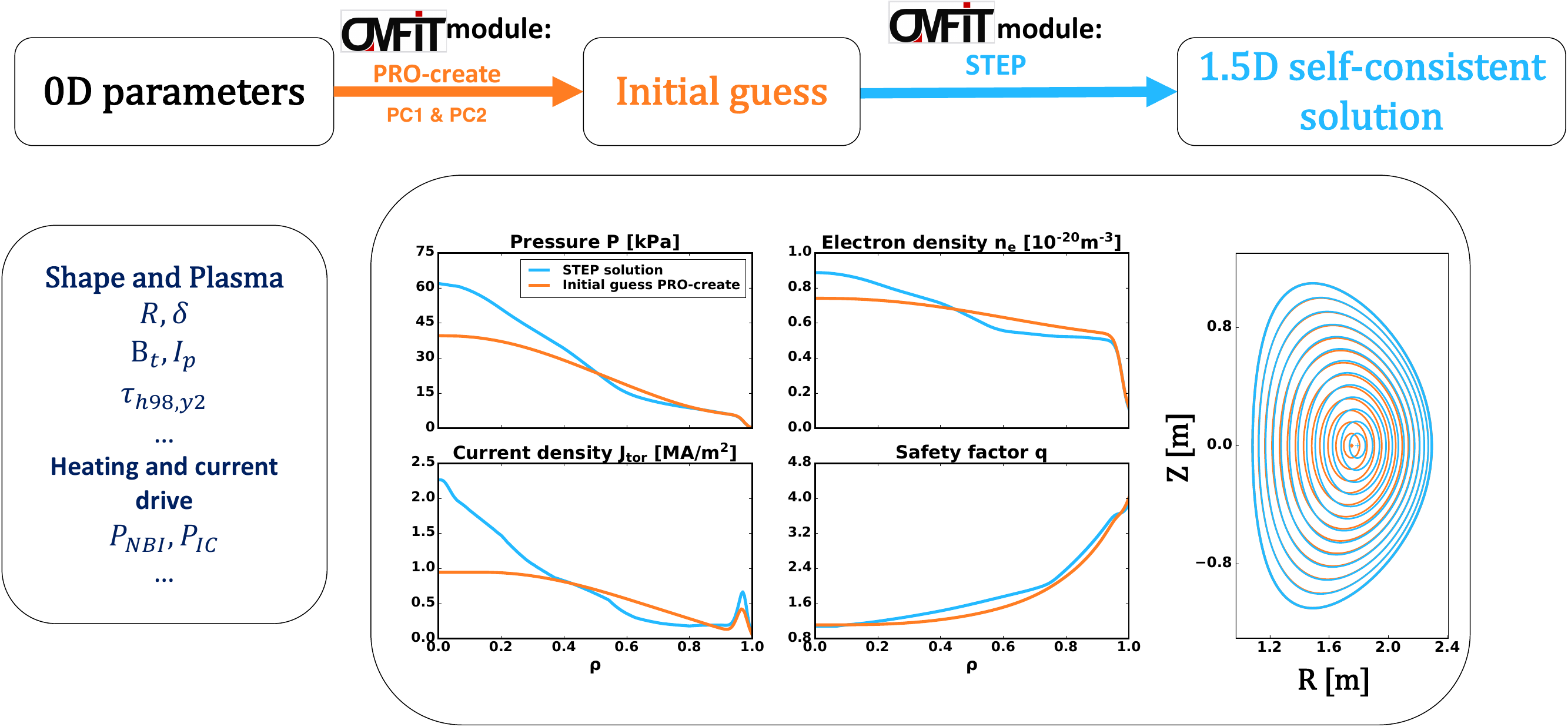}
    \caption{Zero dimensional engineering parameters are transformed by PRO-create (PC1 \& PC2) into an initial guess for the plasma configuration, which are evolved to a self-consistent, stationary physics solution by STEP.}
    \label{fig:zero_to_profile}
\end{figure*}

\emph{Ab initio} 1.5D modeling of tokamak plasmas requires a starting equilibrium solution, kinetic profiles (species densities, temperatures, and rotation), and sources profiles (heating, current drive, and torque). A new module named PRO-create has been created to generate a physically plausible (though not necessarily self-consistent) starting point for 1.5D modeling in the OMFIT framework \cite{meneghini2015integrated}. Figure \ref{fig:zero_to_profile} illustrates the workflow from zero dimensional engineering parameters to the fully self-consistent stationary 1.5D solution. PRO-create has two stages, the first equilibrium stage PC1 is initialized with a simple equilibrium using a constant pressure gradient and a polynomial current density profile, and a second kinetic pressure profile stage PC2, with bootstrap and current drive contributions to the current density profile.

\begin{figure}
    \centering
    \includegraphics[width=0.5\textwidth]{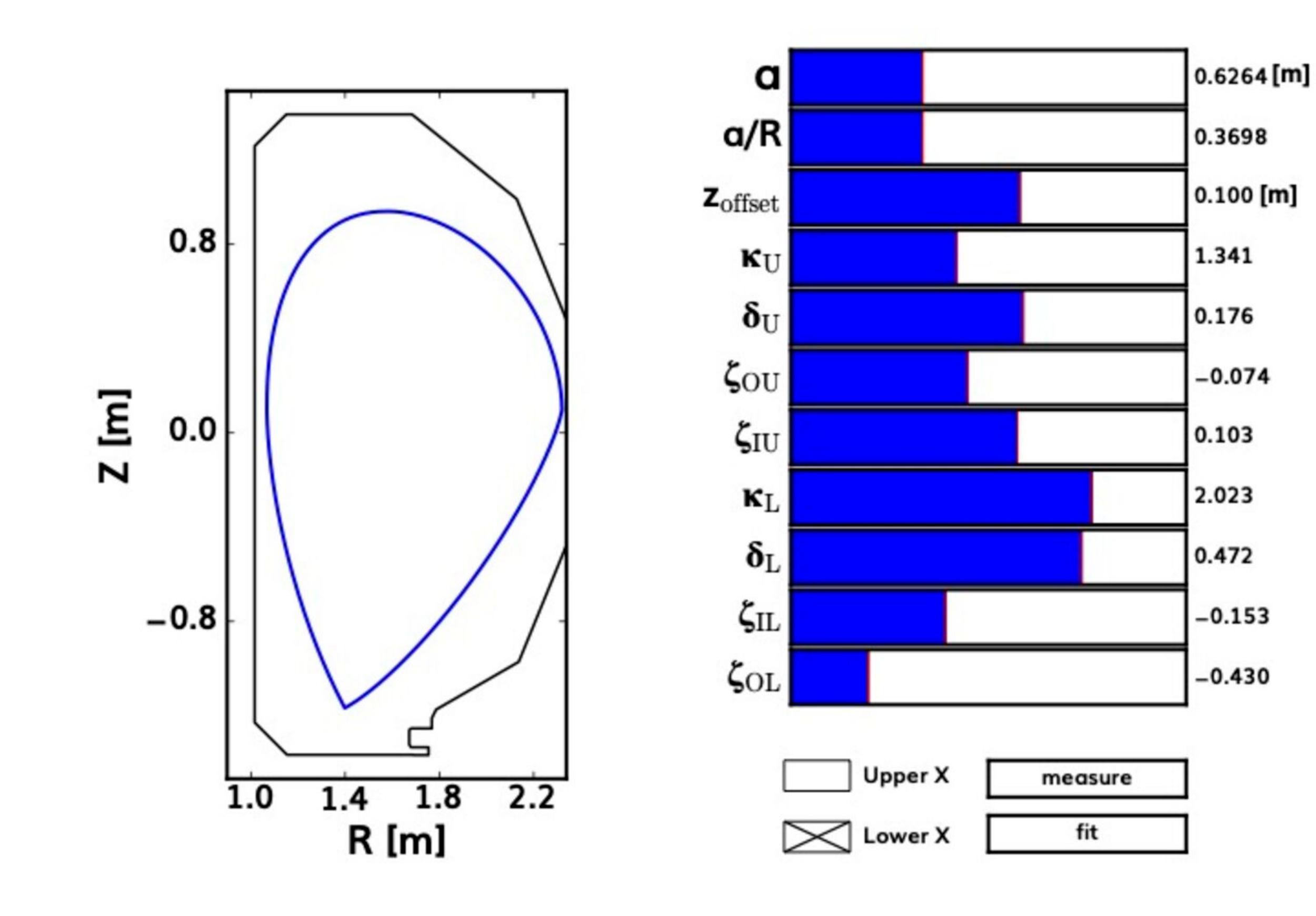}
    \caption{Graphical User Interface of PRO-create's interactive Boundary parameterization.}
    \label{fig:PRO-create}
\end{figure}

In the first stage PC1, PRO-create initializes with the fixed boundary equilibrium solver CHEASE \cite{lutjens1996chease}, assuming a constant pressure gradient $\frac{dP}{d\psi_{N}}$, where $P$ is the pressure and $\psi_{N}$ is the normalized poloidal flux. {\color{black}The plasma boundary can be given as any arbitrary closed boundary}. The graphical user interface (GUI) of PRO-create's parameterization of the fixed boundary is shown in Figure \ref{fig:PRO-create}. The interactive boundary shape tool implemented in PRO-create allows the user to extend the shape to include X-points, up-down asymmetry and squareness. The current density profile term labeled PC1 is a polynomial of the form:
\begin{equation}
\begin{split}
    <J_{\phi}/R>(\psi_{N}) = & c_{I}  \bigg( \underbrace{I_{p} (1-\psi_{\rm N})^\alpha}_{\text{\normalfont PC1 stage}} + \\
    & \underbrace{<J_{BS}/R>(\psi_{N}) + <J_{CD}/R> (\psi_{N})}_{\text{\normalfont PC2 addition}} \bigg)
\end{split}
\label{equation:jtor}
\end{equation}
where $<J_{\phi}/R>$ is the flux surface average of the toroidal current density over the major radius, $I_{p}$ is the total plasma current, $c_{I}$ is a scalar to ensure the integral of averaged $<J_{\phi}/R>$ over the area is equal to the plasma current, and $\alpha$ is a scalar that controls the broadness of the current profile (defaulted to $\alpha = 1.4$, {\color{black} this is only an initial guess and not consequential for the end result see reference \cite{snyder2004elms} for a discussion about the parameter}). In the term labeled PC2, the $J_{BS}$ is calculated with the NEO 2021 adaption of the Sauter formula \cite{redl2021new} and the $<J_{CD}/R>$ term  denotes the contribution of current drive from the auxiliary heating system, and is calculated using equation \ref{equation:jcd}.

In the second stage PC2, H-mode kinetic profiles are then defined on the normalized $\psi_{N}$ grid for the next equilibrium iteration. The initial temperature profiles have the form of a core polynomial and a hyperbolic tangent at the pedestal:
\begin{equation}
    T_{e}(\psi_{\rm N}) = T_{i}(\psi_{\rm N}) = T_{1}(\psi_{\rm N}) + T_{2}(\psi_{\rm N})
    \label{equation:te}
\end{equation}
With $T_{1}$ and $T_{2}$:
\begin{equation}
    T_{1}(\psi_{\rm N}) = \frac{1}{2} c_{t} \left(\tanh \left( \frac{\psi_{\rm N} - x_{\text{ped}}}{w_{\text{ped}}}\right) + \tanh(1) \right) 
\end{equation}

\begin{equation}
    T_{2}(\psi_{\rm N}) = H(1 - \overline{\psi}_{\rm N}) (T_{\text{core}} - T_{1}(\psi_{\rm N}=0)) \left( 1 - \overline{\psi}_{\rm N} ^{a_{\text{in}}} \right) ^{a_{\text{out}}}.
\end{equation}
Here, $H$ the heavy-side step function, $c_{t} = \frac{T_{\text{ped}} - T_{\text{edge}}}{\tanh(1)}$, $T_{\text{core}},T_{\text{ped}}, T_{\text{edge}}$ the core, pedestal, and edge temperatures respectively; $\overline{\psi}_{\rm N} =  \psi_{\rm N} / (x_{\text{ped}} - w_{\text{ped}})$, with $x_{\text{ped}} = 1 - w_{\text{ped}}$, where $w_{\text{ped}}$ is the pedestal width as defined by \cite{snyder2009development} (code that determines the pedestal height that conforms to the peeling and ballooning stability boundary). The core polynomial exponents for the temperature profile are $a_{\text{in}, T}=1.2,$ $a_{\text{out},T}=1.4$, the same profile shape is used for the density but with $a_{\text{in},\,ne}= 1.1$ and $a_{\text{out}, \,ne}=1.1$. {\color{black}These parameters are set to these values to be consistent with the EPED1 parameterization see reference \cite{snyder2004elms}}. The initial pedestal height is obtained from the EPED-NeuralNet \cite{meneghini2017self} (EPED-NN, neural net version of EPED1). The ion densities are set by enforcing quasi neutrality as well as a given $Z_{\text{eff}}$ profile. For a two species system of one fully ionized main impurity species and one fuel species the ion densities are expressed as:

\begin{equation}
    n_{\text{impurity}}(\psi_{N}) = n_{e}(\psi_{N}) \left( \frac{Z_{\text{eff}}(\psi_{N}) - 1}{Z_{\text{impurity}}^{2} - Z_{\text{impurity}}}\right)
\end{equation}

\begin{equation}
    n_{\text{fuel}}(\psi_{N}) = n_{e}(\psi_{N}) - Z_{impurity} \, n_{\text{{impurity}}}(\psi_{N})
\end{equation}
where $n_{\text{impurity}}$ and $z_{\text{imp}}$ are the impurity density and charge state respectively and $n_{\text{fuel}}$ is the fuel species density (hydrogen / deuterium / tritium).

The broad initial guess for the angular frequency (rotation) profile is set as:

\begin{equation}
    \omega_0(\psi_{N}) = 0.9 \omega_{0\,\text{axis}} e^{-3 \psi_{N}} + 0.1 \omega_{0\,\text{axis}}
\label{equation:rot}
\end{equation}
where $\omega_{0\,\text{axis}}$ is the angular frequency (rotation) in rad/s at the magnetic axis.

With the kinetic pressure profile and the complete current density profile described in equation \ref{equation:jtor} a second iteration of CHEASE is taken to find the equilibrium consistent with the kinetic pressure profile and current density profile.

\begin{table*}
\caption{Heating and current drive parameter assumptions}
\begin{ruledtabular}
\begin{tabular}{ccccc}
 Heating system& $f_{cd}$ & $f_{e}=P_{\text{source}}/P_{e}$ & $\text{width}_{e}$ / $\text{width}_{i}$ & $n_{ge}$ / $n_{gi}$\\ \hline
 Neutral Beam Injection & 0.18  & $\text{Sivukhin}(\rho)$ \cite{stix1972heating} & 0.27/0.2 & 1.0/0.7 \\
 Electron Cyclotron Heating  & 0.2  & 1.0 & 0.15 $\frac{0.606[m]}{\text{minor radius}[m]}$ & 1 \\
 Ion Cyclotron Heating  & 0.125 & 0.25 & 0.15 $\frac{0.606[m]}{\text{minor radius}[m]}$  & 1\\
\label{table:hcd}
\end{tabular}
\end{ruledtabular}
\end{table*}

Idealized heating and current drive profiles are implemented for neutral beam injection (NBI), electron cyclotron heating (ECH), ion cyclotron heating (ICH), and lower-hybrid heating (LH) systems each with  different current drive efficiencies and ion-to-electron energy deposition ratio. The heating profiles are defined by equation \ref{equation:qe} as Gaussians of nth order. These heating profiles are required during the flux-matching in the self-consistent evolution in STEP.

\begin{equation}
    q_{e}(\rho) = c_{iq} f_{e}(\rho) P_{\text{source}} \, \, \exp^{\left( \frac{\rho - \rho_{0}}{w_{0}}\right)^{ng_{e}}}
\label{equation:qe}
\end{equation}

\begin{equation}
    q_{i}(\rho) = q_{e}(\rho) (1 - f_{e}(\rho))
\label{equation:qi}
\end{equation}

Here $c_{iq}$ is a volume density scalar to ensure the volume integrated electron energy density source $q_{e}$ is equal to the absorbed power of the heating system $P_{\text{source}}$, $f_{e}$ is the electron fraction of the total power, $w_{0}$ the Gaussian width parameter, $\rho_{0}$ the normalized position of the center of the peak, and $n_{g}$ the order of the Gaussian. The parameters for each heating system can be found in Table \ref{table:hcd}. The width and the Gaussian order $n_g$ of the NBI parameterization were deduced from running ~50 instances of FREYA \cite{goldston1981new} (Neutral Beam deposition code) on DIII-D experimental shots. The width of the ECH and ICH is found by simulation using TORAY\cite{kritz1982ray} (ECH ray tracing code)  and GENRAY \cite{smirnov2001genray} (ray tracing code for ICH and LH) scaled down by minor radius to capture the tokamak size aspect. The current drive of each heating system assumes the same profile shape as the energy density source but with a current drive efficiency $\eta_{CD}$ for each system \cite{kikuchi2012fusion} (chapter 6 of reference [\citen{kikuchi2012fusion}]).

\begin{equation}
    <J_{CD}/R>(\rho) = c_{ij} \eta_{CD} P_{\text{source}} \, \, \exp^{\left( \frac{\rho - \rho_{0}}{w_{0}}\right)^{n_{ge}}},
\label{equation:jcd}
\end{equation}
with $c_{ij}$ an area density scalar to ensure the area integrated current density is equal to the total current drive of the system. The particle source and momentum have the same profile as the NBI heating but with particle source amplitude $S_{NBI} = P_{NBI}/ (e E_{NBI})$ with $P_{NBI}$ the absorbed heating power of the NBI and $E_{NBI}$ the energy of the beam particles and for the total injected momentum $T_{NBI} = S_{NBI} \sqrt{\frac{2 e E_{NBI}}{m_{beam}}}m_{beam} \sin{\theta_{V}} $ with $\theta_{V}$ the vertical tilt angle of the NBI with respect to the plasma cross section and $m_{beam}$ the mass of the injected NBI particles.

The output of PRO-create is stored in the IMAS data schema (IDS) \cite{imbeaux2015design} format using the OMAS implementation \cite{meneghini2020neural}, the equilibrium information is stored in the \emph{equilibrium} IDS, the kinetic profiles in \emph{core profiles} and the heating and current drive in the \emph{core sources} IDS.

\subsection{Self-consistent theory based solution with STEP}
\begin{figure}
\includegraphics[width=0.5\textwidth]{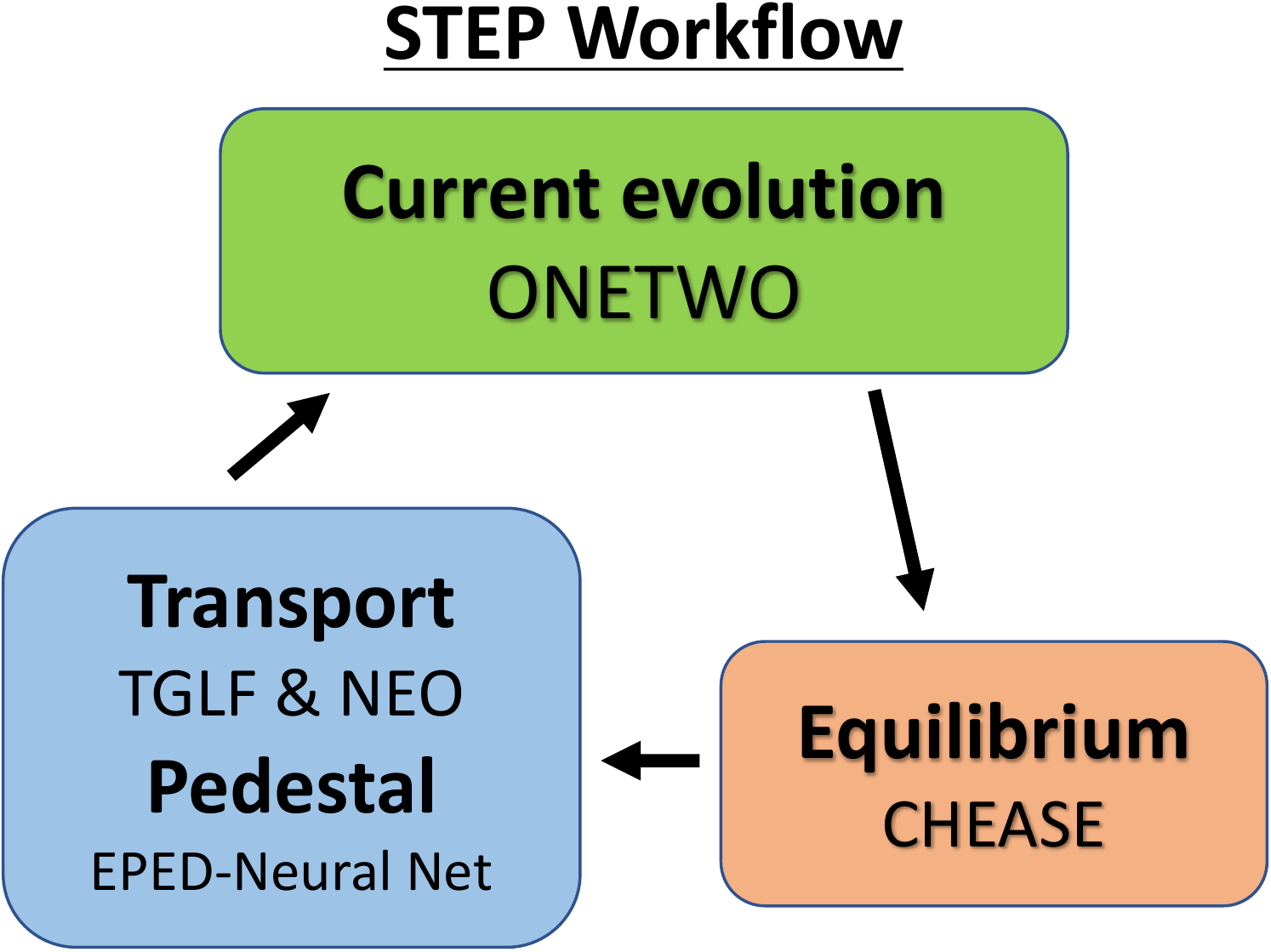}
\caption{Specific STEP workflow cycle used in this paper with the following steps: ONETWO:current evolution, CHEASE: equilibrium, TGYRO: pedestal and transport.}
\label{fig:stepwf}
\end{figure}

\begin{figure}
\includegraphics[width=0.5\textwidth]{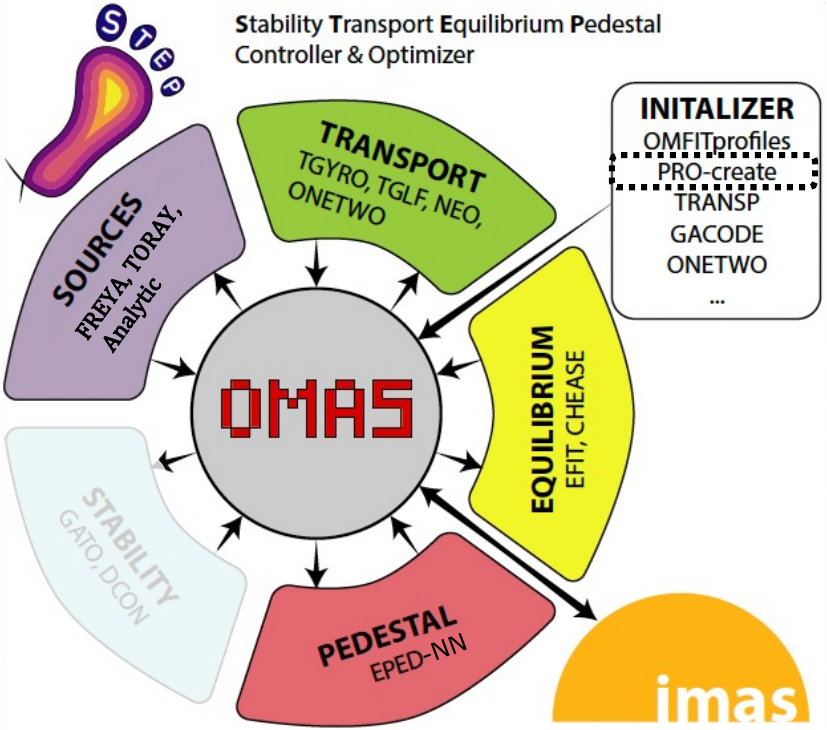}
\caption{Data flow of the STEP workflow, initializing with PRO-create to form a starting point in the OMAS data structure implementation where each physics step reads and writes data to the same centralized data structure (adapted from reference \cite{meneghini2020neural}).}
\label{fig:stepdf}
\end{figure}

The 1.5D solution provided by PRO-create is physically plausible but not yet consistent so it is used as the starting point for the STEP \cite{lyons:2022} workflow which iterates over core sources, transport, pedestal and equilibrium codes (as illustrated in Figure \ref{fig:stepwf}) until it converges on a self-consistent solution. Each physics step reads and writes data to and from the same centralized data structure (as illustrated in Figure \ref{fig:stepdf}). The physics steps used in this paper are summed up below:
\begin{itemize}
    \item For the equilibrium step the fixed boundary code CHEASE  \cite{lutjens1996chease} is used, CHEASE makes the pressure and current profiles self-consistent based on the solution of the grad-shavranov equation using the kinetic $P(\rho)$ and $<J_{\phi}/R>(\rho)$ as inputs in the \emph{equilibrium} IDS. The solution provides the reference magnetic geometry used to map profiles. An anomalous resistivity was added to keep to keep the safety factor above one, essentially emulating the sawtoothing process; nevertheless, the influence of sawteeth on temperature was not considered in this approach.
    \item For the current evolution, ONETWO \cite{pfeiffer1980onetwo} is used which calculates the self-consistent ohmic current profile and bootstrap current profile. ONETWO  is setup to evolve the current till steady state, this assumes the shot has undergone a few current diffusion time steps from the start of the flattop. 
    \item In order to make the density, temperature and rotation profiles consistent with the fluxes according to validated \cite{2020APS..DPPZ04007N} reduced models of neoclassical and turbulent (gyrokinetic) fluxes TGYRO \cite{candy2009tokamak} is used. TGYRO calls TGLF SAT2 \cite{staebler2020geometry} with electromagnetic contributions (EM) and NEO \cite{belli2008kinetic} for the turbulent and neoclassical fluxes respectively. The particle, energy and momentum source is gathered from the \emph{core sources} IDS to which TGYRO adds the synchrotron and line radiation as well as the bremsstrahlung. The transport grid is set with 8 points evenly spaced from $0.2 \leq \rho \leq 0.8$ with the boundary condition $\rho =0.8$, this boundary point is set as the solution of EPED-NN \cite{meneghini2017self} which calculates the width and the height of the pedestal. The pedestal height is blended with the boundary condition at $\rho =0.8$ see reference \cite{meneghini2016integrated} for more details. The solutions of the fluxes are stored in the \emph{core transport} IDS.
\end{itemize}

In the TGYRO step the line-averaged density is matched to the experimental or target value by shifting the pedestal up and down to match the synthetic interferometer (this is done after each converged TGYRO solution). Each tokamak has its own interferometer path, for consistency the synthetic interferometer's path was chosen as a horizontal path through the magnetic axis.

In Table \ref{table:evolution} an overview can be seen of the most physically relevant plasma parameters and profiles that are used and calculated in PRO-create and STEP.

\begin{table*}
\caption{Overview of plasma parameters and profiles used in PRO-create and STEP}
\begin{ruledtabular}
\begin{tabular}{cccc}
 Quantity&Evolved or Static& Description  \\ \hline
 $I_{p}$ & static : 0D Input & Total toroidal plasma current \\
 $B_{t0}$ & static : 0D Input & Toroidal magnetic field on-axis \\
 $\kappa$& static: 0D Input & Elongation \\
 $\delta$ & static: 0D Input  & Miller up down symmetric triangularity \\
 $Z_{\text{eff}}$& static: 0D Input & Effective charge Z for all ion species \\
 $n_{e,\text{line}}$ & static : 0D Input & Line averaged density \\
  $P_{\text{source}}, S_{\text{source}}$ &static: 0D Input  & Total absorbed power and total particle source  \\
 $q_{e}(\rho)$ & static : Equation \ref{equation:qe} & Electron heating energy density \\
 $q_{i}(\rho)$ & static : Equation \ref{equation:qi} & Ion heating energy density \\
$n_{e, \text{sep}}$ & static : $n_{e \, ped} / 4$  & Separatrix density \\
$q_{e, ohmic}$ & evolved : ONETWO & Ohmic heating profile \\
$n_{e, ped}$ & evolved : Synth diagnostic & Pedestal density, EPED definition \cite{snyder2009development} \\
$n_{e}(\rho)$ &evolved : TGYRO  & Electron density profile \\
$T_{e}(\rho), T_{i}(\rho)$ &evolved : TGYRO &  Electron and Ion temperature \\
$\omega_{0}(\rho)$ &evolved : TGYRO & Angular frequency (rotation) \\
$<J_{tor}/R>(\rho)$ & evolved : ONETWO & Flux surface average of the toroidal current density over R \\
$\Psi(R,Z)$ &evolved : CHEASE & 2D flux surfaces map \\
$\Psi_{\text{bound}}$ &static : Miller parameterization & 2D fixed boundary \\
\label{table:evolution}
\end{tabular}
\end{ruledtabular}
\end{table*}
Predictions of plasma pulses using this integrated modeling workflow starting with very limited information requires us to make a number of assumptions, yet we will show that it yields good predictions for global confinement when staying within the validated parameter space of underlying physical models and assumptions.

\section{Results}

\subsection{Sample DIII-D case from experiment}

\begin{table*}
\caption{0D input parameters for DIII-D discharge 81499 at 4000 ms}
\begin{ruledtabular}
\begin{tabular}{ccccccccccc}
 $I_{P}[MA]$ & $B_{t}[T]$ & $Z_{\text{eff}}$ & $n_{e,\text{line}}[10^{20}\,m^{-3}]$ & $R_{\text{major}}[m]$ & $a_{\text{minor}
}[m]$  &  $\kappa$ & $\delta$ & $P_{NBI}[MW]$ & $E_{NBI}[keV]$  \\ \hline
 1.34 & 1.91 & 2.32 & 0.48 & 1.69 & 0.63 & 1.68  & 0.32 & 5.74 & 72.6 & 
\label{table:0dparams}
\end{tabular}
\end{ruledtabular}
\end{table*}

\begin{figure}
    \centering
    \includegraphics[width=0.5\textwidth]{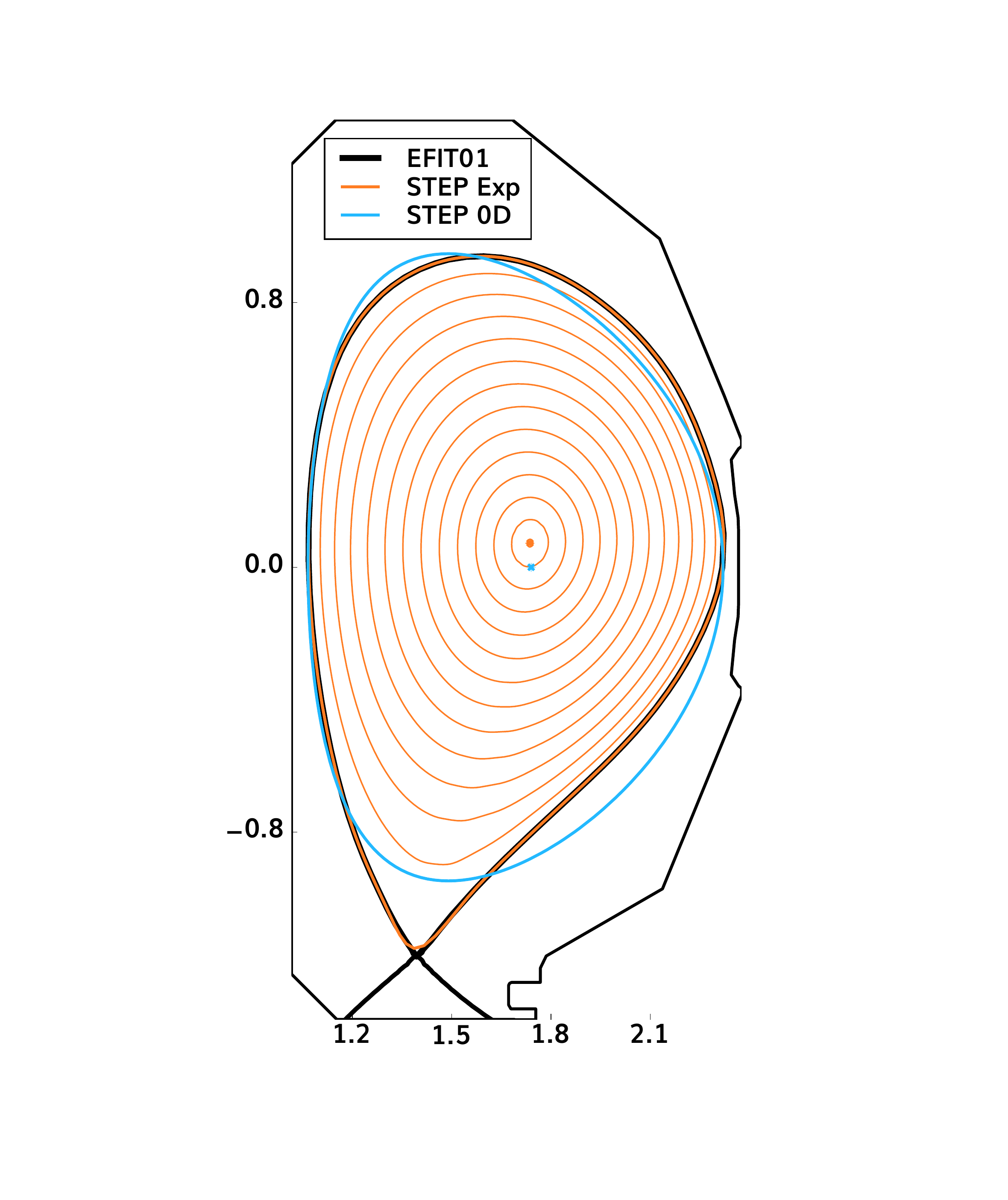}
    \caption{Plasma equilibrium of shot 81499 at 4000 ms for the experimental EFIT01 in black, STEP-experimental in orange and STEP-0D in blue.}
    \label{fig:Equilibrium_shot}
\end{figure}

\begin{figure}
    \centering
    \includegraphics[width=0.5\textwidth]{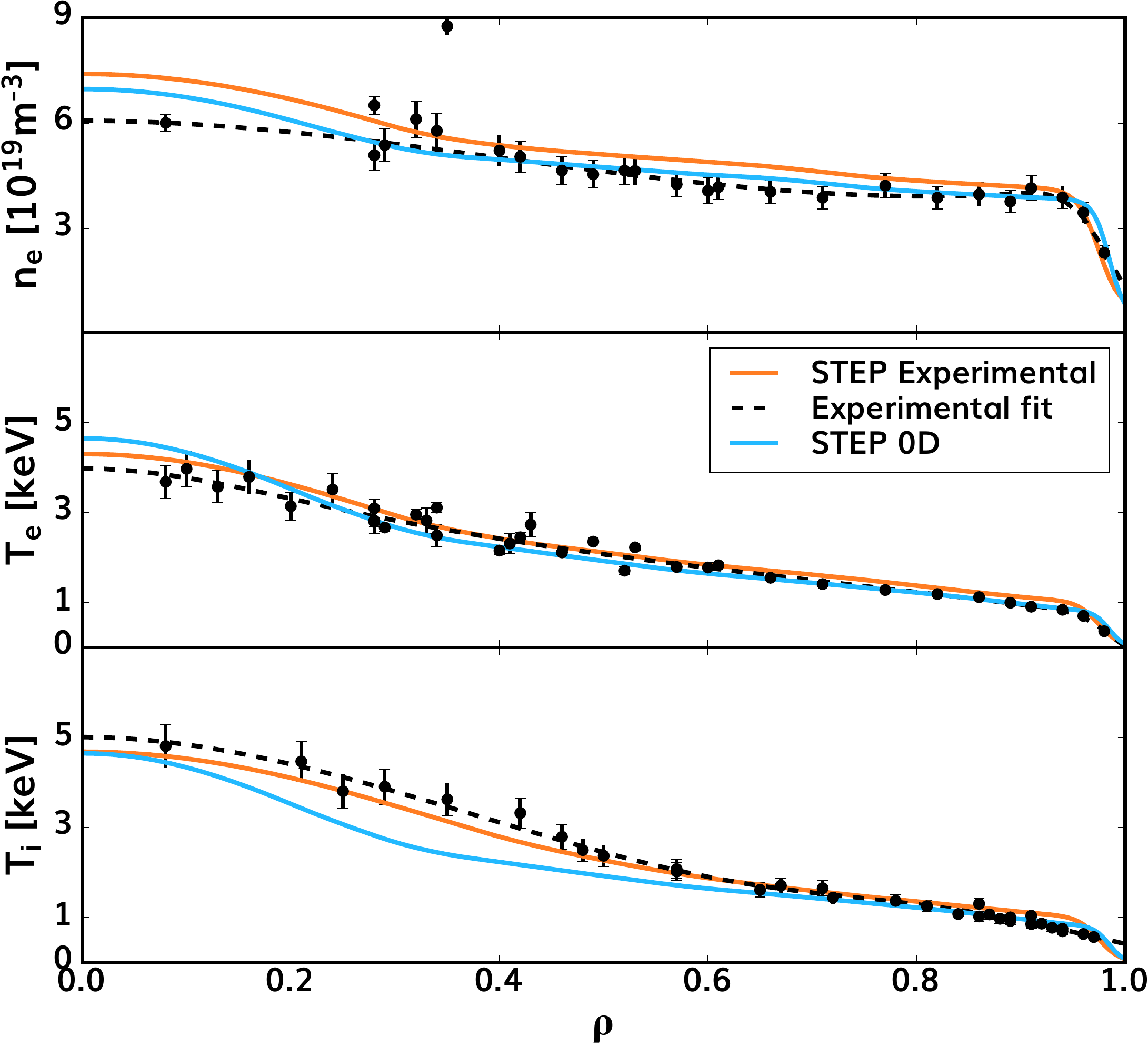}
    \caption{Core profile comparison between experiment (black), STEP-experimental (orange) and  STEP-0D (blue) for DIII-D shot 81499 at 4000 ms, (top) electron density, (middle) electron temperature and (bottom) ion temperature.}
    \label{fig:Profile_db_shot}
\end{figure}

The PRO-create/STEP workflow has been validated on a number of DIII-D discharges. Shot \# 81499 is shown as an example of this validation set. Information about this shot is publicly available \cite{roach20082008}. This shot is a conventional H-mode with global parameters as described in Table \ref{table:0dparams}. Information about this shot includes plasma profiles like the density, rotation and temperature profiles but also heating and current drive profiles from interpretative analysis in addition to the EFIT01 \cite{lao2005mhd} plasma boundary (EFIT01 is a magnetic only reconstruction of the equilibrium using EFIT and is valid for the plasma boundary shape). Setting up STEP from the experimental equilibrium including experimental heating and current profiles, is referred to as STEP-experimental. This approach is compared to setting up STEP based on 0D quantities, which is referred to as STEP-0D. The ITPA global H-mode database records the plasma boundary in up-down symmetric Miller geometry while the experiment is a lower single-null plasma with up-down asymmetric geometry. To reconcile the difference in plasma volume between STEP-0D and STEP-experiment the symmetric Miller elongation of STEP-0D is changed to match the recorded plasma volume from the database using a simple optimization algorithm. This is important when comparing volume based quantities like the thermal stored energy. The pedestal model EPED and TGLF both utilize the symmetric Miller parameterization, thereby neglecting to account for the asymmetry caused by the x-point. The equilibrium shape of shot \# 81499 from STEP-0D with adjusted elongation compared to the EFIT01 and STEP-experimental can be seen in Figure \ref{fig:Equilibrium_shot}. The predicted plasma density and temperatures of STEP-0D and STEP-experimental can be seen in Figure \ref{fig:Profile_db_shot}. The electron density and temperature profiles are nearly identical with some slight difference in the ion temperature gradient scale length around $0.3 < \rho < 0.5$. This discrepancy can be explained by contrasting the analytical heating and current drive applied in STEP-0D with the actual heating profiles in the STEP-experimental run; specifically, the former demonstrates a higher electron-to-ion heating ratio, while the latter exhibits more off-axis heating. The matching pedestal density indicates that matching the experimental line averaged density from the experiments using the synthetic diagnostic in the TGYRO step works as intended. 

\subsection{Global H-mode confinement database}
To extend the validation of the STEP modeling beyond single experimental use cases STEP-0D integrated modeling has been applied to a subset of the ITPA global H-mode database. This database contains data from multiple machines with confinement times ranging from milliseconds to 1200 milliseconds. The database was filtered to exclude elongation smaller than 1.4 ($\kappa > 1.4$ as the EPED-NN training's data does not contain small elongation), filtering shots that do not have deuterium as main fuel species (to exclude Hydrogen and Helium plasmas) and filtering plasmas that did not record triangularity. 

\begin{figure}
    \centering
    \includegraphics[width=0.5\textwidth]{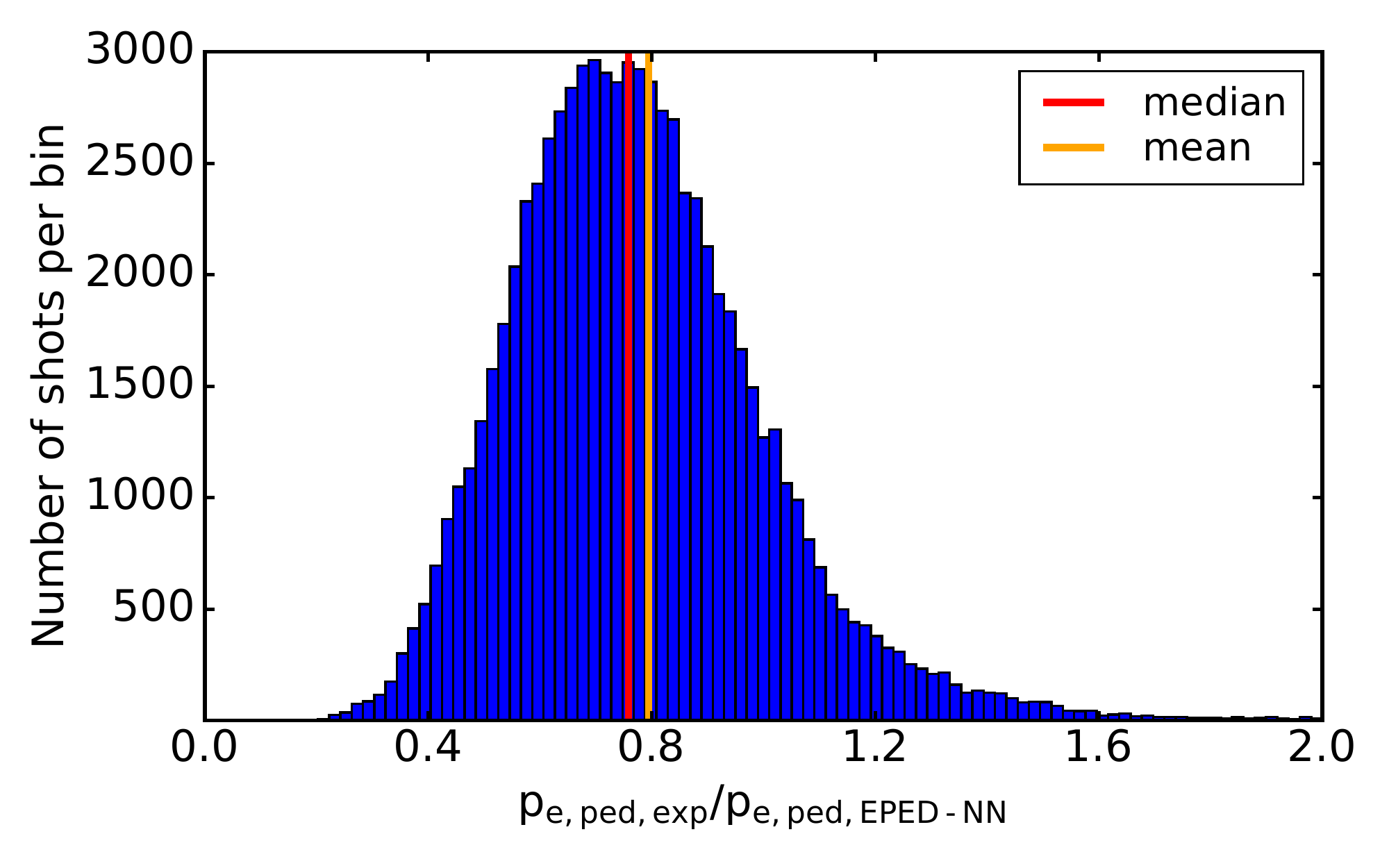}
    \caption{The histogram shows the ratio between the experimental pedestal electron pressure and the predicted electron pedestal using EPED-NN. The mean and median of the ratio are 0.794 and 0.758, respectively. The 76728 experimental data points were obtained by automated tanh function fitting to Thomson scattering measurements in real space for 655 DIII-D shots.}
    \label{fig:eped_nn_vs_exp}
\end{figure}

This subset contains 600 discharges from 7 different tokamaks with carbon walls and divertor, with an exception of CMOD which has a molybdenum wall (fully ionized Neon was used to model the CMOD impurities). For all the cases in the database the PRO-create and STEP settings were unchanged and the STEP workflow steps through the cycle described in Figure \ref{fig:stepwf} three times. Although PRO-create and STEP support up-down asymmetric plasma boundaries {\color{black}the data recorded in the ITPA H-mode database includes only the up-down symmetric Miller quantities}. The same approach as in Section II was used by matching the total volume by changing the elongation slightly. The database includes the line averaged density but no information about the pedestal or separatrix density therefore PRO-create sets an initial guess for these quantities as fractions of the line averaged density, $n_{e \, \text{ped}} = n_{e\,\text{line}} / 1.3$, $n_{e\,\text{sep}} = n_{e \, \text{ped}} / 4$. A flat profile is prescribed, $Z_{\text{eff}}(\rho) = Z_{\text{eff}}$ where $Z_{\text{eff}}$ comes from the database. If this value is missing $Z_{\text{eff}}= 2$ is chosen. The initial on axis rotation is set as $\omega_{0 \text{axis}} = 10^{5}\, [\text{rad}/\text{s}]$ with an exception for shots without any NBI. For those the rotation is not evolved and set to zero. The heating and current drive profiles are updated throughout the run with the analytic form as presented in section A heating and current drive codes weren't used as there is not enough information in the database to set-up each heating system so core deposition at $\rho_{0}=0.0$ was assumed for all cases. EPED predicts the peak pressure of the ELM cycle, it is unknown at what point in the cycle the data is taken therefore the pedestal height is reduced to an effective 80 \% of the EPED(-NN) prediction. This reduction is supported by experimental evidence as can be seen in Figure \ref{fig:eped_nn_vs_exp} where 76728 EPED-NN predictions were made and compared to automated tanh function fitting to Thomson scattering measurements in real space for 655 DIII-D shots where the mean ratio of experimental to predicted sits at {$p_{\text{e, ped, exp}}/p_{\text{e, ped, EPED-NN}} = 0.794$}.

\begin{figure}
    \centering
    \includegraphics[width=0.49\textwidth]{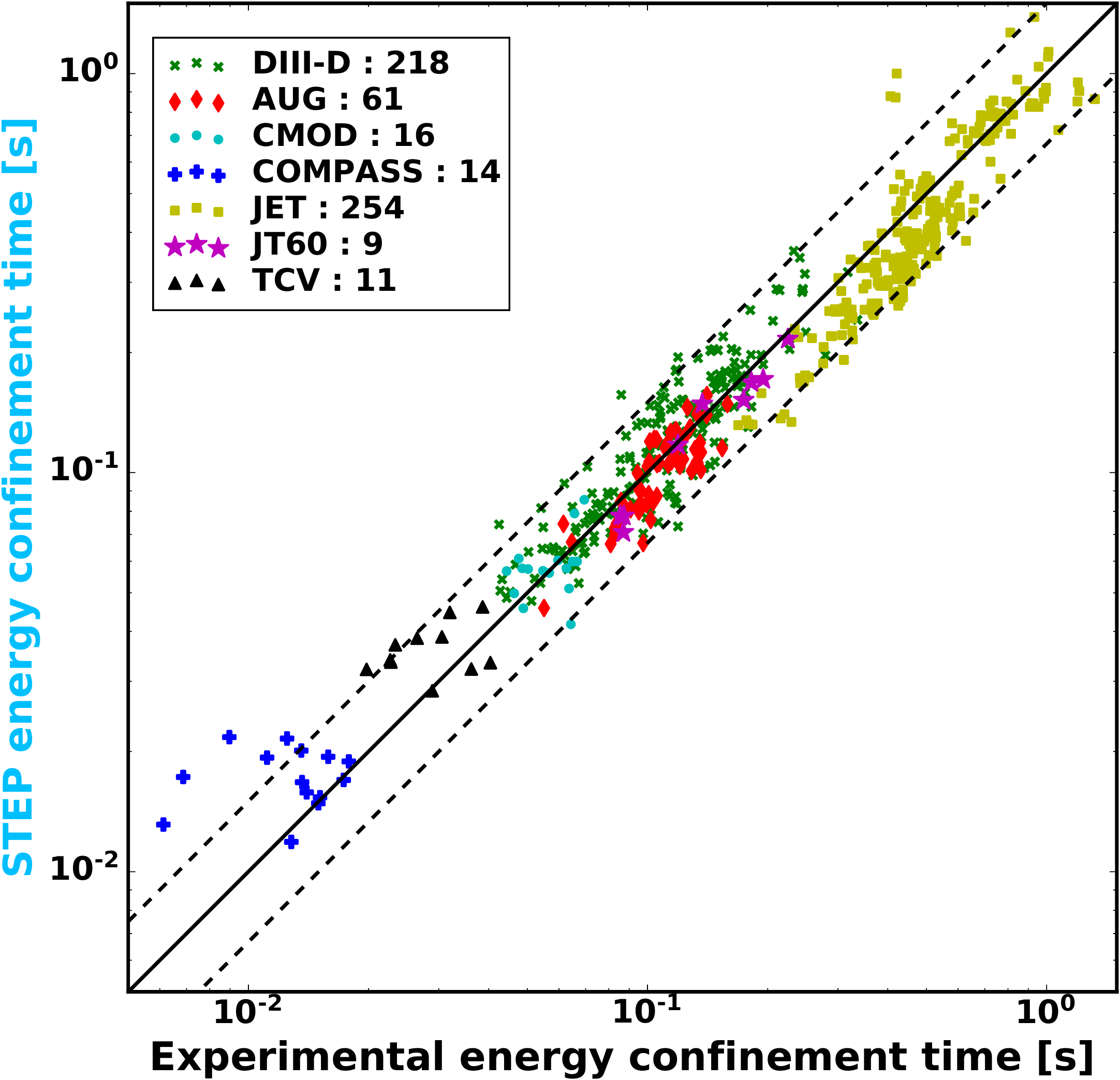}
    \caption{Experimental confinement time compared to the STEP-0D result for the subsection of the ITPA database, the black dotted lines denote a 50\% error from the experimental energy confinement time, the mean relative error of the STEP-0D results to the experiments for this subset is 19\%.}
    \label{fig:STEP_v_tau}
\end{figure}

\begin{figure}
    \centering
    \includegraphics[width=0.45\textwidth]{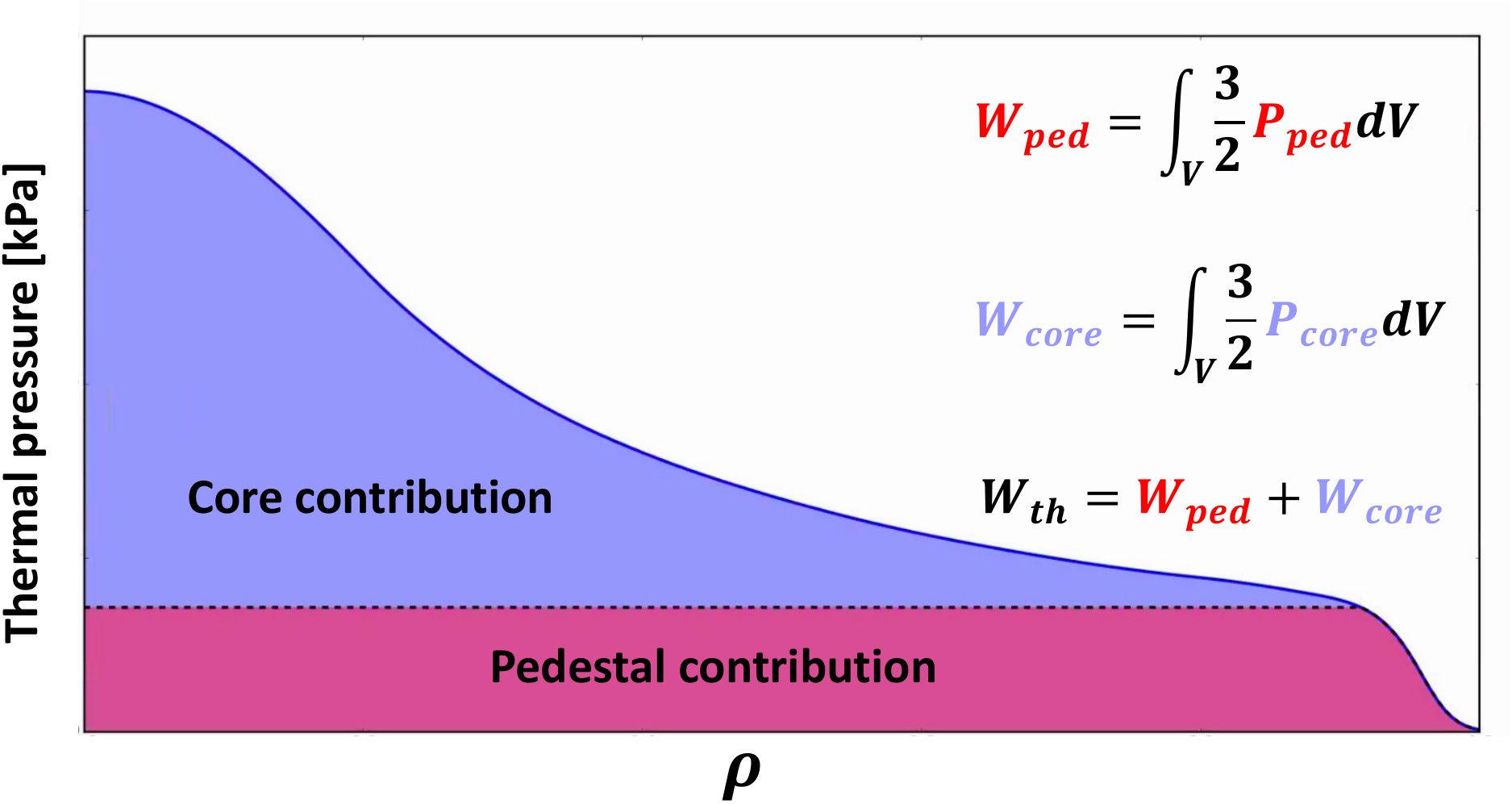}
    \caption{Visualization of the Pedestal (magenta) and core contribution (purple) to the thermal stored energy.}
    \label{fig:ped_to_core}
\end{figure}

\begin{figure}
    \centering
    \includegraphics[width=0.45\textwidth]{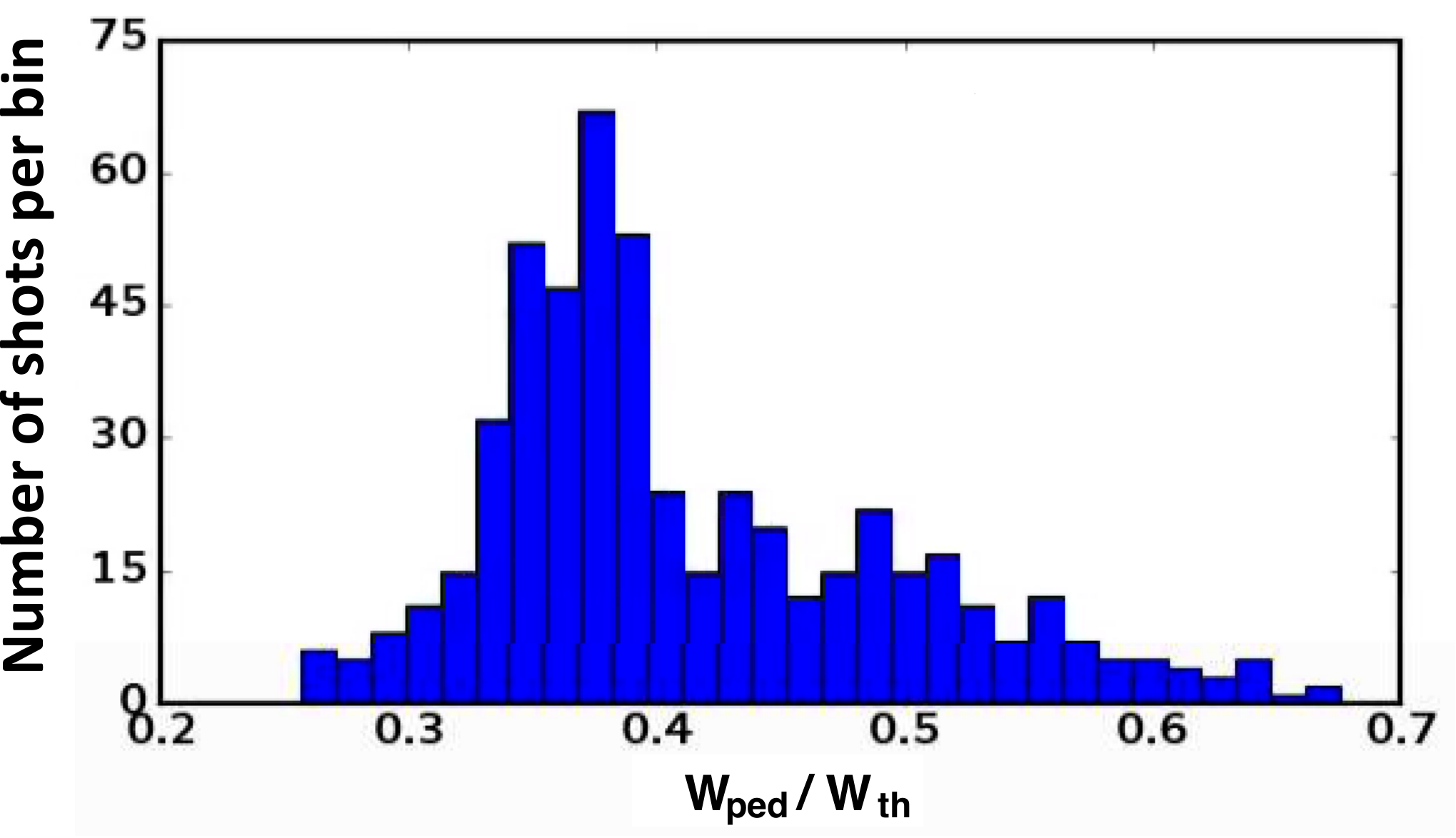}
    \caption{Histogram of the ratio of total thermal stored energy to the pedestal stored energy for the STEP results of the validation database subset.}
    \label{fig:wth_wped}
\end{figure}

To compare STEP-0D to the experiments, the experimental energy confinement time is plotted against the STEP-0D simulated energy confinement time as a regression plot in Figure \ref{fig:STEP_v_tau}. The mean relative error of the STEP-0D results to the experiments for this subset is 19\%, for context the IPB98(y,2) scaling law has a mean relative error of 22\% for the same subset. The discharges from AUG, CMOD, DIII-D JET and JT60 follow the $\tau_{E, \text{exp}} = \frac{W_{th, \text{exp}}}{P_{lth, \text{exp}}} = \tau_{E, \text{STEP}} = \frac{W_{th, \text{STEP}}}{P_{lth, \text{exp}}}$ line (with $P_{lth, \text{exp}}$ the experimentally defined loss of power of the thermal species)  with some scatter. The three COMPASS experimental points with the shortest confinement times received a high amount of injected power. Both the EPED and EPED-NN models predict significantly higher pedestal pressures (6-12 kPa) in these plasmas compared to the predictions made by Komm et al. \cite{komm2017contribution}. One plausible explanation for the difference in $\tau_e$ between the experimental results and the STEP predictions is that the pedestals in the experimental plasmas were constrained by smaller Type III ELMs \cite{sartori2004study} driven by ballooning instabilities, which are not accounted for in EPED. To support this claim the importance of the pedestal height on the confinement time in the pedestal stored energy ($W_{ped}$) component is compared to the core stored energy ($W_{core}$) component by splitting them up, as is illustrated in Figure \ref{fig:ped_to_core}. Although the core gradients and pedestal height are not entirely independent, the pedestal height plays a significant role in determining both the total stored energy and the height of the core profiles. The ratio of total stored thermal energy ($W_{th}$) to the pedestal contribution  can be seen in Figure \ref{fig:wth_wped}, which concludes that the pedestal contribution is significant for conventional type I ELMy H-mode shots.

\begin{figure}
\includegraphics[scale=0.8]{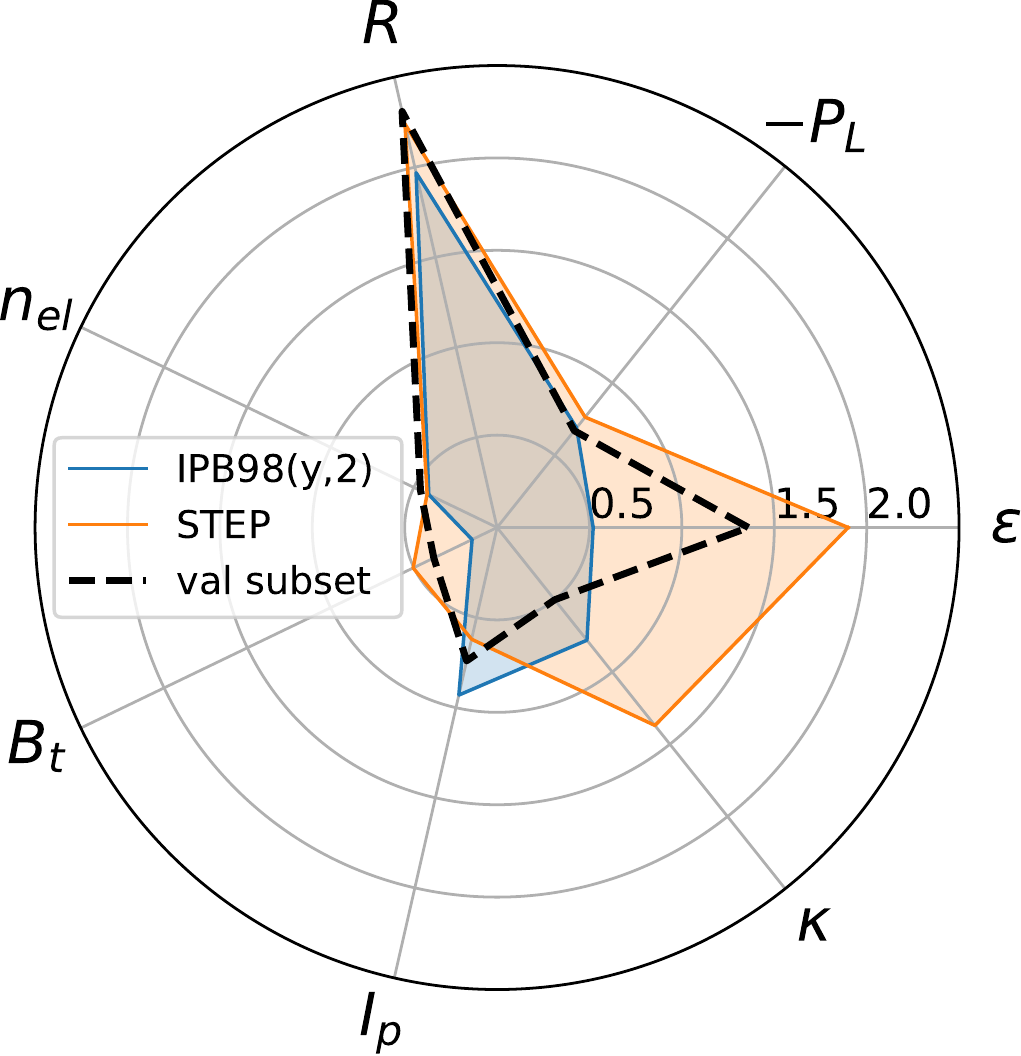}
\caption{Radar plot showing the exponents of weighted linear regressions for the scaling law tau IPB98(y,2) (blue), the validation subset (black) and the STEP-0D result (orange), all exponents are a positive contributions with the exception of the loss power.}\label{fig:radar}
\end{figure}

Currently there is no equivalent pedestal model for type III ELM shots as well as for ELM free operational regimes like negative triangularity, EDA-H, grassy ELM regime. Exploration of ELM-free discharges \cite{paz2021plasma} is important for fusion power plant sized devices as ELM heat fluxes scale unfavorably in fusion power plant parameter spaces. The formulation of such models is therefore critical in performing any predictive simulation and scoping of parameter spaces for fusion power plant operation. STEP would prove a useful tool for coupling these new type of edge models to benchmark against experiment and improve the models. 

\begin{figure}
    \centering
    \includegraphics[width=0.5\textwidth]{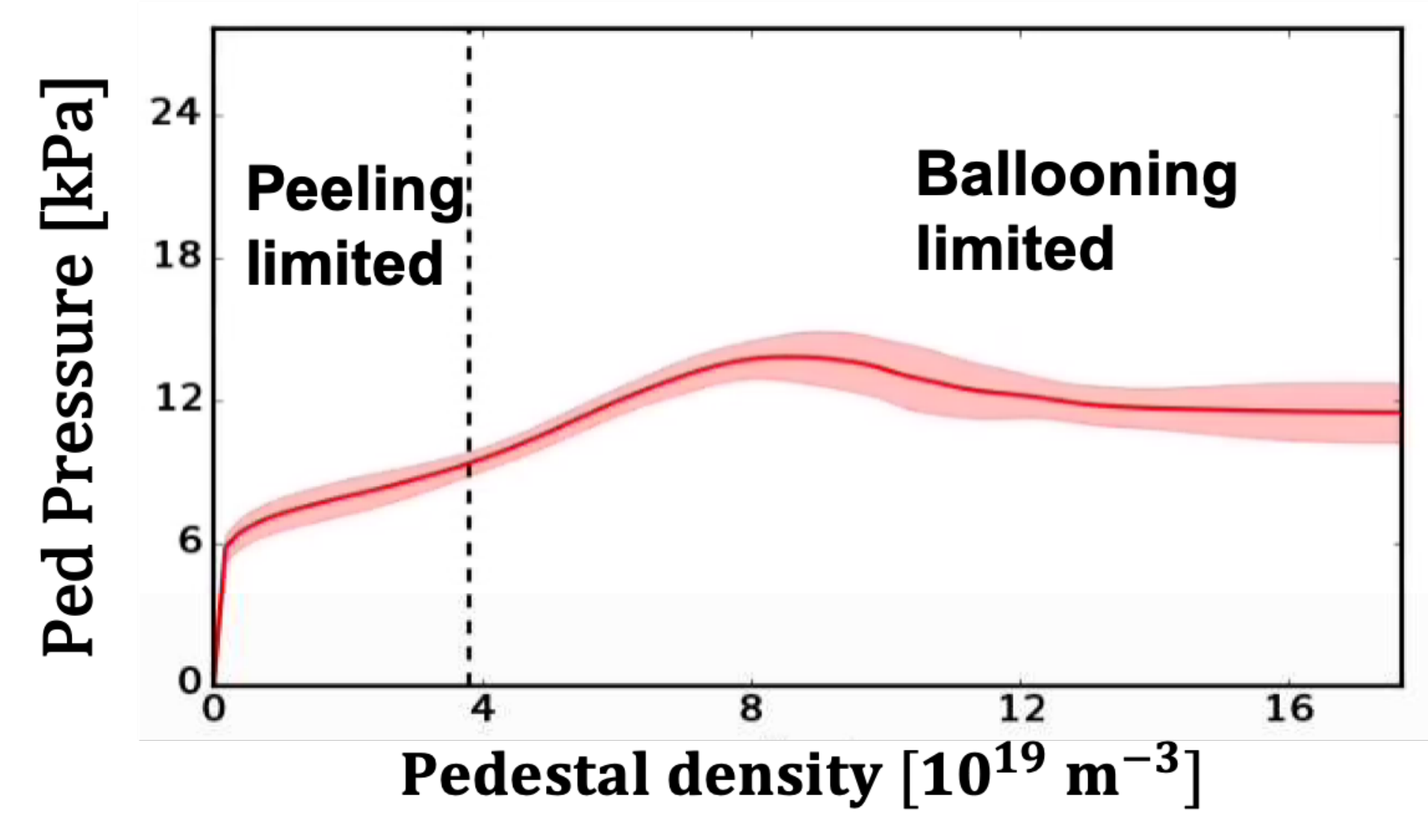}
    \caption{EPED neural net sensitivity scan, showing the pedestal pressure as a function of the pedestal density where the EPED-NN input parameters are kept constant except for the pedestal density for a DIII-D nominal operation point (dashed lines), the shaded region in red denotes the uncertainty of the model.}
    \label{fig:sensitivity}
\end{figure}

Weighted least squares regression (WLS) of the experimental database and STEP results can be made to infer differences between the IPB98(y,2) scaling law and modeling results. The entire database used to generate the IPB98(y,2) scaling consists of ~10 000 shots, its WLS exponents can be seen in blue in Figure \ref{fig:radar} where the weight is a function of the number of plasma discharges in the database $w_{i} = 1 / (2+ \text{Int}(\sqrt{N_{i}}/4)) $(as described in McDonald et al. \cite{mcdonald2007recent}). A WLS on the subset used in this study gives the exponents in black. The best fitting exponents for the WLS regression ($R^2 = 0.998$) of the STEP results are shown in orange in Figure \ref{fig:radar}. The exponents of the WLS regression of STEP indicate agreement with the data subset with the exception to elongation $\kappa$ and the inverse aspect ratio $\epsilon$. Emphasis should not be given to the exact exponents of the STEP power law, as for example sensitivity analysis of the pedestal density (Figure \ref{fig:sensitivity}) on the pedestal pressure indicates that the underlying physical behavior is not power-law like but depends on the operating regime. Peeling limited pedestal pressures scale inversely with pedestal density in the ballooning limited regime whereas peeling limited regimes scale with the pedestal density. Additionally it is difficult to attribute what causes the difference in power law exponents like elongation and inverse aspect ratio as they are heavily correlated to the other parameters. Similar analysis but then for L-mode plasmas is done see reference \cite{angioni2023dependence}. This successful validation of the PRO-create STEP workflow for ELMy H-mode plasmas gives confidence in predicting the fusion power output of fusion power plant sized devices. Utilization of this workflow finds discrepancies between 0D-system code predicted fusion power and the predicted fusion power (see reference [\citen{lyons:2022}] for an application of this). Another ongoing effort of this STEP - PRO-create workflow is the design of a plasma scenario for a new tokamak device called BEST (Burning plasma Experimental Reactor). 

\subsection{Confinement predictions for fusion reactors}
\begin{figure}
\includegraphics[width=0.5\textwidth]{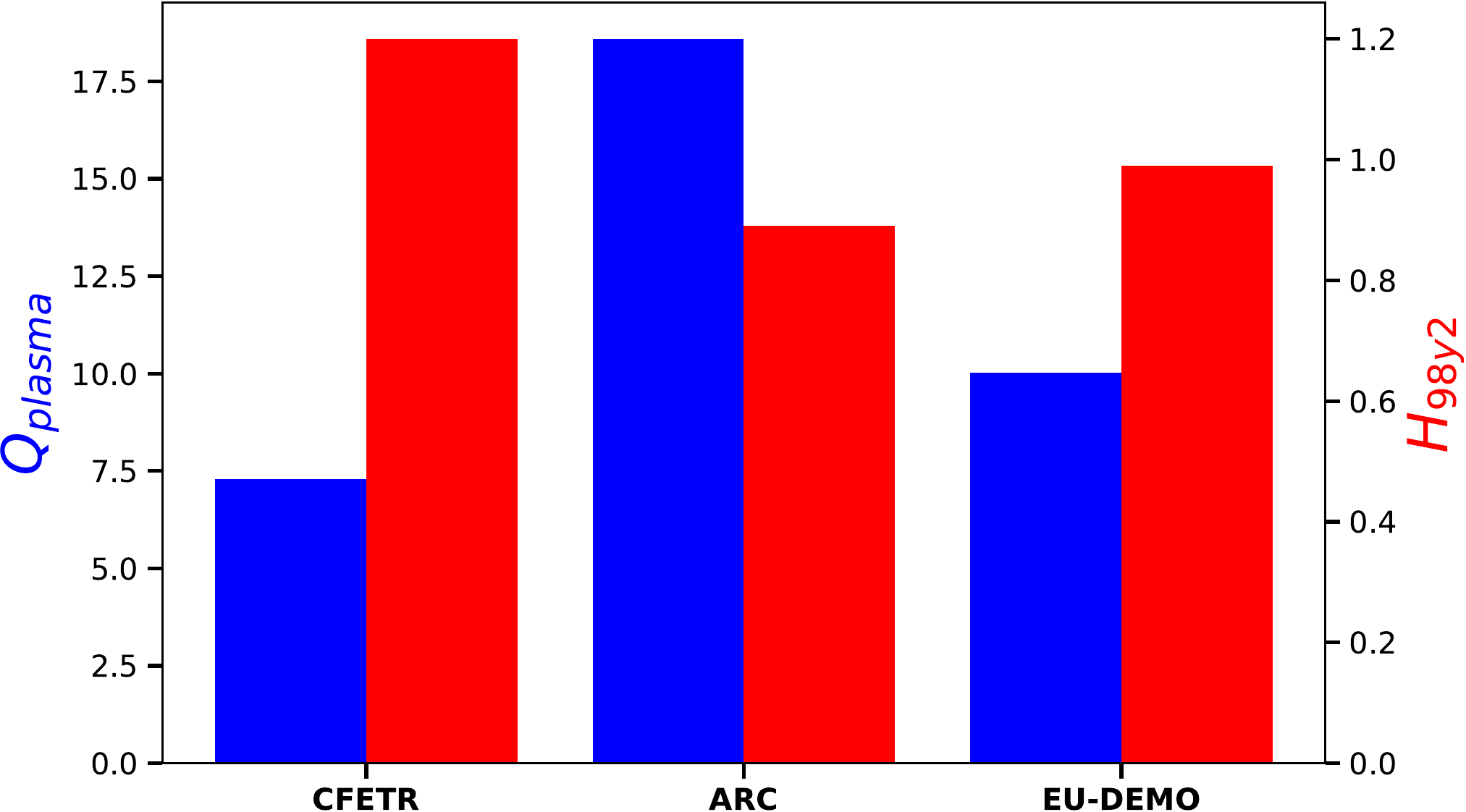}
\caption{Fusion reactor plasma performance results from the STEP-0D simulations with in blue $Q_{\text{plasma}} = P_{\text{fusion}}/P_{\text{aux}}$ and $H_{98y2} = \tau_{E, \text{STEP}} / \tau_{E, \text{IPB98y2}}$ in red. The EU-DEMO case is with the increased auxiliary heating to 80 MW and the ARC case is at the result at the peeling ballooning limit.}
\label{fig:reactor_perf}
\end{figure}

With the validated STEP workflow starting from 0D parameters predictions can be made for proposed fusion reactors. To demonstrate this capability predictions were made for CFETR (Table 1 baseline case  \cite{chen2021integrated}), ARC (ARC RPL mode Table 2 right most column \cite{frank2022radiative}) and EU-DEMO (Table 1 EU-DEMO 2018 \cite{siccinio2022development}). The 0D parameters of each case are summarized in table \ref{table:0dparams_reactors}. The plasma rotation profile $ \omega_{0}(\rho) $ for all cases was set to zero due to the large size device in the case of CFETR and EU-DEMO and for ARC zero torque from H\&CD systems. A Helium specie was included in all cases as 0.03 fraction of electron density. The same settings for each of the physics codes were used (see section B), EPED was ran instead of EPED-NN as these cases lie outside the NN trainings dataset. For TGLF the SAT 2 model was used including electromagnetic contributions where TGYRO evolved the density and temperatures while keeping the pedestal density fixed. The predicted plasma performance and confinement factors can be seen in Figure \ref{fig:reactor_perf}. 

\begin{table*}
\caption{0D input parameters for CFETR, ARC, EU-DEMO where --> denotes changes to the reference parameters.}
\begin{ruledtabular}
\begin{tabular}{c|cccccccccc}
 Device & $I_{P}[MA]$ & $B_{t}[T]$ & $Z_{\text{eff}}$ & $n_{e,\text{ped}}[10^{20}\,m^{-3}]$ & $R_{\text{major}}[m]$ & $a_{\text{minor}
}[m]$  &  $\kappa$ & $\delta$ & $ P_{aux}[MW] $ & ion species \\ \hline
 CFETR & 13.0 & 6.5 & 2.2 & 0.7 & 7.2 & 2.2 & 2.01  & 0.43 & 80 & D,T,He,Ar\\
 ARC & 14 & 11.5 & 1.13 & 2.5 & 4.2 & 1.2 & 1.62 & 0.35 & 22.7 & D,T,He,Kr\\

 EU-DEMO & 17.75 & 5.86 & 2.19 & 0.45 & 9.07 & 2.92 & 1.65 & 0.33 & 50 --> 80 & D,T,He,Kr
\label{table:0dparams_reactors}
\end{tabular}
\end{ruledtabular}
\end{table*}
\subsubsection*{ARC}

In the ARC reference, the simulation simulates a lower-hybrid heating of 22.7 MW, employing the same heating profiles utilized in the STEP analysis. This was done by setting the transport solver's boundary point at $\rho=0.9$ with $T_{i} = T_{e}$ = 2.5 keV. which was derived through scaling L-mode plasmas from CMOD. This method was in line with the accurate replication of the L-mode scaling law, as reported by Angioni in 2023 \cite{angioni2023dependence}. However, this assumption may be overly optimistic for an L-mode since analysis with EPED suggests that the peeling-ballooning limit is at 147 kPa at the pedestal (220 kPa at $\rho=0.9$). Furthermore, the resulting fusion power from the simulation is particularly sensitive to the transport boundary condition at $\rho=0.9$. This sensitivity is apparent in the results presented in Table \ref{table:fusionpowers}. Here, the fusion power in H-mode at the peeling-ballooning limit was comparable to that in the reference scenario. Yet, when considering 80\% or less of the peeling-ballooning limit pressure, the profiles collapse due to the loss of fusion heating and the highly radiative krypton species. The detailed transport analysis of the L-mode edge is thus crucial to ascertain the validity of the reference scenario, considering the high sensitivity of the fusion power to the boundary condition. Moreover, the H-mode peeling-ballooning limit might be overly optimistic, given that the plasma power falls far short of the threshold required for the L to H-mode transition scaling.

\begin{table}
\caption{Sensitivity of the fusion power to the transport boundary condition at $\rho = 0.9$ where the H-mode peeling ballooning (PB) limit sits at 224 kPa and reference at 200 kPa.}
\begin{ruledtabular}
\begin{tabular}{c|cccc}
 cases  & PB-limit & Reference & 80\% of PB-limit\\ \hline
 P($\rho =0.9$) [kPa] & 224 & 200 & 181 \\ \hline
Fusion power [MW] & 420 & 290 & 2.3 
\label{table:fusionpowers}
\end{tabular}
\end{ruledtabular}
\end{table}

\subsubsection*{CFETR}
The CFETR case employs 30 MW of NBI with a beam particle energy of 1 MeV, coupled with 50 MW of ECH. The NBI deposition is at $\rho_{0} = 0.0$ and the EC deposition is at $\rho_{0} = 0.3$. Under these conditions, the workflow yields a fusion power output of 583 MW. A higher fusion power output of 952 MW, as suggested by the reference, is attainable under a hybrid operation scenario. This scenario is predicated on optimizing the plasma current profile, which enhances overall plasma performance. While it is feasible to conduct similar optimizations to reach this power output, such measures fall beyond the scope of the current study.
\subsubsection*{EU-DEMO}
In the 2018 scenario for EU-DEMO, a specific heating and current drive system was not selected, leading to the adoption of 30 MW of NBI with a beam particle energy of 1 MeV. Concurrently, 20 MW of ECH was implemented with deposition positions at $\rho_{0} = 0.0$ and $\rho_{0} = 0.3$ respectively. The pedestal density was assumed to be 80\% of the pedestal top as specified by the case. The reference fusion power was projected by a 0D system code to be approximately 2012 MW. However, when 50 MW of auxiliary heating was applied, the workflow only predicted a fusion power of 550MW, indicating an under-powered system. An increase in auxiliary power to 80 MW, with 50 MW from NBI and 30 MW from ECH, resulted in a modest increase in fusion power to 802 MW. This is notably less than the expectations set by the 0D design study cited as a reference. While further optimization is beyond the scope of the current study, the gap to achieve 2000 MW of fusion power appears substantial, suggesting the necessity of substantial revisions or improvements.

\section{Conclusions}
A new STEP workflow has been developed that predicts plasma profiles and equilibrium starting from 0D tokamak engineering parameters.
This workflow consists of generating physically plausible plasma profiles consistent with the plasma equilibrium using PRO-create followed by iteration between physics models for the equilibrium, sources, core transport, and pedestal to obtain a self-consistent solution using STEP. Systematic validation against the ITPA global H-mode confinement database has shown that STEP-0D can predict the energy confinement time with an average mean relative error (MRE) of less than 19\%. The successful validation of the PRO-create STEP workflow for ELMy H-mode plasmas gives confidence in predicting the fusion power output of fusion power plants. Predictions of this workflow were made on CFETR, ARC and EU-DEMO where Conventional H-modes with moderate H-factors between 0.9
and 1.2 were found. The conventional methodology employed in the design of fusion power plants (FPP) starts with the identification of a operating point that is found by running a 0D system codes optimization. The chosen design is then further refined using more sophisticated, high-fidelity models. However, this method proves to be inefficient at times as the results of high-fidelity models do not always coincide with the optimal solutions identified by the 0D system codes. This discrepancy becomes particularly evident when the design deviates from the assumptions underpinning the system code's scalings and the reduced set of equations, such as those valid for a given aspect ratio or operating regime. This paper presents a viable alternative: the ability to obtain self-consistent 1.5D physics-based stationary solutions that start from the same 0D input parameters utilized by the system codes. We envision that this capability, when integrated with engineering and costing models in an optimization workflow, could form the foundation for a novel FPP design workflow. This new workflow would eliminate the need for iterating between plasma models of varying fidelity and thus avoid wasting time reconciling inevitable discrepancies. Further development of the physics capabilities of this work involves coupling different pedestal models to include regimes of ELM-free operation and developing and validating a robust workflow for advanced tokamak scenarios (that rely on large radius internal transport barriers to achieve high confinement and steady-state operation).

\subsection*{ACKNOWLEDGEMENTS}
This material is based upon work supported by the U.S. Department of Energy, Office of Science, Office of
Fusion Energy Sciences, using the DIII-D National Fusion Facility (DE-FC02-04ER54698), a DOE Office of Science user facility, under
Awards DE-SC0017992 (AToM) and DE-FG02-95ER54309  (GA Theory Grant).
Part of the data analysis was performed using the OMFIT integrated modeling framework \cite{meneghini2015integrated}. The Authors thank Cihan Akcay for editorial suggestions.

\subsection*{DISCLAIMER}
This report was prepared as an account of work sponsored by an agency of the United States Government.
Neither the United States Government nor any agency thereof, nor any of their employees, makes any
warranty, express or implied, or assumes any legal liability or responsibility for the accuracy, completeness,
or usefulness of any information, apparatus, product, or process disclosed, or represents that its use would not
infringe privately owned rights. Reference herein to any specific commercial product, process, or service by
trade name, trademark, manufacturer, or otherwise, does not necessarily constitute or imply its endorsement,
recommendation, or favoring by the United States Government or any agency thereof. The views and opinions
of authors expressed herein do not necessarily state or reflect those of the United States Government or any
agency thereof.

\nocite{*}
\bibliography{aipsamp}

\end{document}